# Quantum secure direct communication with quantum memory

Wei Zhang[1,3], Dong-Sheng Ding[1,3*], Yu-Bo Sheng[2#], Lan Zhou[2], Bao-Sen Shi[1,3†] and Guang-Can Guo[1,3]

[1]*Key Laboratory of Quantum Information, Chinese Academy of Sciences, University of Science and Technology of China, Hefei, Anhui 230026, China*

[2]*Key Laboratory of Broadband Wireless Communication and Sensor Network Technology, Nanjing University of Posts and Telecommunications, Ministry of Education, Nanjing 210003, China*

[3]*Synergetic Innovation Center of Quantum Information and Quantum Physics, University of Science and Technology of China, Hefei, Anhui 230026, China*

Corresponding authors: * *dds@ustc.edu.cn*
# *shengyb@njupt.edu.cn*
† *drshi@ustc.edu.cn*

Quantum communication provides an absolute security advantage, and it has been widely developed over the past 30 years. As an important branch of quantum communication, quantum secure direct communication (QSDC) promotes high security and instantaneousness in communication through directly transmitting messages over a quantum channel. The full implementation of a quantum protocol always requires the ability to control the transfer of a message effectively in the time domain; thus, it is essential to combine QSDC with quantum memory to accomplish the communication task. In this paper, we report the experimental demonstration of QSDC with state-of-the-art atomic quantum memory for the first time in principle. We used the polarization degrees of freedom of photons as the information carrier, and the fidelity of entanglement decoding was verified as approximately 90%. Our work completes a fundamental step toward practical QSDC and demonstrates a potential application for long-distance quantum communication in a quantum network.

The importance of information and communication security is increasing rapidly as the Internet becomes indispensable in modern society. Quantum communication exploits unconventional quantum properties to provide unconditional security and novel ways of communicating. There are many modes of quantum communication, such as quantum key distribution (QKD) [1-5], quantum secret sharing [6, 7], quantum teleportation [8-12], and quantum secure direct communication (QSDC) [13-16], which have been widely explored over the past 30 years. In particular, some exciting developments, including entanglement-based QKD over 144 km [3], and quantum teleportation and entanglement distribution over 100 km free-space channels [11], have been achieved that have laid the foundation for future long-distance quantum communication and quantum networks.

As a branch of quantum communication, QSDC can transmit secret messages over a quantum channel directly without setting up a private key session. In this protocol, the sender and receiver share the prearranged entangled photon pair first, thus setting up the communication channel. After the receiver obtains one photon of the pair, the sender encodes the remaining photon with one of the four unitary operations, $I$, $\sigma_z, \sigma_x$, and $\sigma_{iy}$, which correspond to the encoding information 00, 01, 10, and 11, respectively. Finally, the receiver performs a Bell-state measurement for photons to decode the information after receiving the second photon. For efficiency, practical QSDC is preferable to the multiplexing scheme, which means that the sender and receiver share $N$ pairs of entangled photons in the time domain or space domain. During this process, some pairs of photons are chosen randomly for a channel security check in each distribution. Clearly, this protocol eliminates key management, which is a potential security loophole, and ciphertext. Additionally, it greatly promotes security and instantaneousness in communication [13-16].

The first QSDC protocol exploited the properties of Bell states and used a block transmission technique [13]. In 2003, the standard criterion for QSDC was explicitly clarified [14], for which a two-step QSDC protocol using the Einstein–Podolsky–Rosen pair block was proposed. Additionally, QSDC can also be used to achieve QKD, and it has a higher capacity than typical QKD [13, 17]. Furthermore, a QSDC protocol based on single photons was also proposed [15], and it is easier to achieve now because of the rapid advance in single-photon devices [18]. Recently, researchers have experimentally demonstrated that QSDC with single photons can work in a noisy environment using frequency coding [16].

Because the full implementation of a quantum protocol always requires the ability to control the transfer of a message effectively in the time domain [19, 20], it is essential to combine QSDC with quantum memory to accomplish the communication task. Currently, an optical fiber delay line is used as a substitute for quantum memory in QSDC to store the encoded photons [16]. Such an approach has played important and helpful roles in the proof-of-principle experimental demonstration. However, for a practical application, a quantum memory is greatly needed because it is robust against decoherence and convertible to quantum states of light on demand [21-26], which indicates effective control for transferring a message in the time domain, therefore a delay line is only suitable for a fixed time-delay situation, but using a quantum memory can provide a more flexible, effective and instantaneous communication, and it is also a necessary element in future quantum network [19]. Using genuine quantum memory is an experimental goal and a challenging task because it requires the coherent storage of entangled single photons, and precise and effective control of quantum states. In this paper, we report the first experimental proof-of-principle demonstration of QSDC with genuine quantum memory. This is a key advance in secure communication based on QSDC.

First, we prepared a hybrid atom-photon entangled state in one atomic ensemble, and then we distributed the photon and stored it in another atomic ensemble, thus establishing memory-memory entanglement. The encoding operation was performed on the retrieved photon from the first atomic ensemble using a dense coding approach [27, 28], and the decoding operation was obtained through density matrix reconstruction after retrieving the stored photon from the second atomic ensemble. We achieved high fidelity of approximately 90% for entanglement decoding for this QSDC experiment.

We briefly describe the basic procedure of QSDC based on the polarization entanglement [13, 14] of photons. Suppose Alice wants to send a message directly to Bob. The detailed steps are as follows:

(1) Alice first prepares $N$ pairs of Bell states $|\phi^+\rangle=(|H\rangle|H\rangle+|V\rangle|V\rangle)/\sqrt{2}$. We assume that $|\phi^\pm\rangle$, and $|\psi^\pm\rangle$ are the four polarized Bell states: $|\phi^\pm\rangle=(|H\rangle|H\rangle\pm|V\rangle|V\rangle)/\sqrt{2}$ and $|\psi^\pm\rangle=(|H\rangle|V\rangle\pm|V\rangle|H\rangle)/\sqrt{2}$, where $|H\rangle$ and $|V\rangle$ are the horizontal and vertical polarized photon states, respectively. Among these $N$ pairs, Alice choses some randomly as check pairs.

(2) Alice and Bob agree that $|\phi^+\rangle$, $|\phi^-\rangle$, $|\psi^+\rangle$, and $|\psi^-\rangle$ encode the bit values 00, 01, 10, and 11 respectively. Alice distributes one photon from each pair of Bell states $|\phi^+\rangle$ to Bob, and hence, sets up the entanglement channel.

(3) After a channel security check, Alice encodes her remaining photons with messages. Alice

performs one of four unitary operations, $I$, $\sigma_z$, $\sigma_x$, or $\sigma_{iy}$, to transform state $|\phi^+\rangle$ to $|\phi^+\rangle$, $|\phi^-\rangle$, $|\psi^+\rangle$, or $|\psi^-\rangle$, respectively. These operations correspond to the encoding information 00, 01, 10, and 11, respectively.

(4) Alice sends her encoded photons to Bob. After Bob receives the photons, he performs the Bell-state measurement to decode the information from Alice, and performs another channel security check simultaneously.

In QSDC, because the transmission of the *N* photons and encoding operations all require some time, the photon pairs shared by Alice and Bob are stored for some time and the storage time is larger than $T_o+L/c$, where $T_o$ is the operation time for Alice and $L/c$ is the transmission time for a photon, where *L* is the communication distance and *c* is the speed of light.

The experimental setup is summarized in Figure 1. The medium used to generate the entanglement was an optically thick ensemble of $^{85}$Rb atoms trapped in a two-dimensional magneto-optical trap (MOT) [29]. A Signal-1 single photon at a 795-nm wavelength entangled with atomic spin waves in MOT A was created with the aid of a beam displacer (BD) after the illumination of the Pump-1 light (30-ns pulse), and then delivered to the second atomic ensemble in MOT B for storage. BD3 and BD4 were used to guarantee the same memory efficiency for different polarized states of Signal 1. With the shutting down of the coupling light, the Signal-1 photon was stored in MOT B as an atomic spin wave, thus establishing the entanglement between spin waves in two atomic ensembles. In this case, light-memory entanglement was stored as memory-memory entanglement. After 50-ns of storage in MOT A, the spin wave was retrieved as a Signal-2 photon, which was obviously entangled with an atomic spin wave in MOT B. Using two half-wave plates (HWPs), we encoded the Signal-2 photon. Then, the Signal-2 photon was delivered to the side where MOT B was located. After that, the atomic spin wave in MOT B is retrieved with a total of 120-ns storage time. Both Signal 2 and the retrieved Signal 1 were detected through a projection measurement to reconstruct the entangled state.

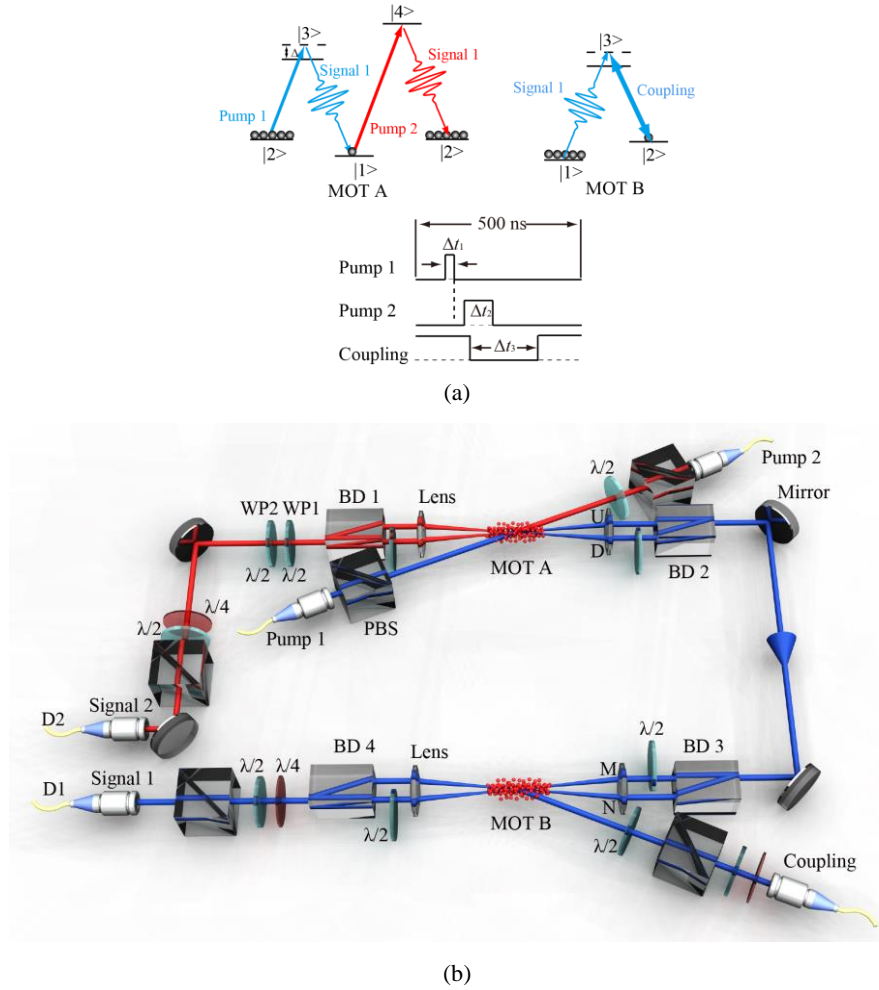

Figure 1 Quantum secure direct communication with atomic ensembles. (a) Energy diagram and time sequence. Pump 1 and Pump 2 are pulses with duration $\Delta t_1$=30 ns and $\Delta t_2$=200 ns, respectively. The time delay set for the spin wave in MOT A is 50 ns and the storage time is set to $\Delta t_3$=120 ns for the spin wave in MOT B. $\Delta$, which represents single photon detuning, is set to +70 MHz, $|1\rangle=|5S_{1/2},F=2\rangle$, $|2\rangle=|5S_{1/2},F=3\rangle$, $|3\rangle=|5P_{1/2},F=3\rangle$, and $|4\rangle=|5P_{3/2},F=3\rangle$. (b) Simplified experimental setup. PBS: polarizing beam splitter; $\lambda/2$: half-wave plate; $\lambda/4$: quarter-wave plate; BD: beam displacer; $U$ $(D/M/N)$: the path; D1/D2: single photon detectors (avalanche diode, Perkin-Elmer SPCM-AQR-15-FC); and WP1/WP2: half-wave plate for coding the Signal 2 photon. Pump 1 is incident obliquely onto the plane consisting of path $U$ and path $D$, with the same angles (1.5 °) to them, and Pump 2 is collinear backward with Pump 1. The coupling light is also obliquely onto plane consisting of path $M$ and path $N$, with the same angle 1.5 °. The power of Pump 1, Pump 2, and the coupling light are 0.2 mW, 4 mW, and 24 mW, respectively.

Our system worked periodically with a cycle time 10 ms, which included 8.7-ms trapping and initial state preparation time, and 1.3-ms operation time for 2,600 cycles, with a cycle time of 500 ns. In each cycle, Pump 1, Pump 2, and the coupling light were pulsed by an acousto-optic modulator; they were all Gaussian beams with a waist of 2 mm. The optical depth (OD) of the atomic ensemble in MOT A and MOT B was approximately 20 and 50, respectively. The coupling efficiency for both paths $M$ and $N$ of Signal 1 from space to fiber was 75%. Signal-1 photons were filtered using three homemade cavities with temperature control, with 45% transmittance and 70 dB isolation. Signal-2 photons were filtered using two homemade cavities with 65% transmittance and 40 dB isolation.

Entanglement, including path-polarization entanglement between the spin wave and photonic polarization, was directly generated with the illumination of Pump 1 through the spontaneous Raman scattering (SRS) process. Because of the conservation of momentum in the SRS process, the initial system had zero momentum, thus the resulting joint state of Signal 1 and the spin wave had zero momentum in K-vector space, hence the spin wave in MOT A entangled with the Signal-1 photon, which can be written as (unnormalized)

$$|\psi_0\rangle = |D_A\rangle|H_{S1}\rangle + e^{i\theta_1}|U_A\rangle|V_{S1}\rangle,$$

where $|D_A\rangle$ and $|U_A\rangle$ denote the spin wave related to the paths $D$ and $U$ in MOT A accordingly, $|H_{S1}\rangle$ and $|V_{S1}\rangle$ represent the generated horizontal and vertical polarizations of the Signal-1 photon, respectively, and $\theta_1$ is the phase difference between paths $D$ and $U$, which was set to zero in this experiment.

After the SRS process, with the aid of the first Mach–Zehnder interferometer in MOT A containing BD1 and BD2 and two half-wave plates, the generated Signal-1 single photon entangled with the spin wave in MOT A was delivered to the MOT B for storage. Using this technique, the entanglement between two atomic ensembles was established as

$$|\psi_1\rangle = |D_A\rangle|N_B\rangle + |U_A\rangle|M_B\rangle,$$

where $|N_B\rangle$ and $|M_B\rangle$ denote the spin wave related to paths $N$ and $M$ in MOT B.

After 50-ns storage in MOT A, we switched on the Pump 2 light; thus, the spin wave in MOT A was retrieved as a Signal-2 photon. In this case, the entanglement between the spin waves in two atomic ensembles was transformed into an entanglement between a Signal-2 photon and spin wave in MOT B:

$$|\psi_2\rangle = |H_{S2}\rangle|N_B\rangle + |V_{S2}\rangle|M_B\rangle.$$

Then, the Signal-2 photon was coded by WP1 and WP2, and then delivered to Bob's site and detected by D1. Additionally, after a total of 120-ns storage, the spin wave in MOT B was retrieved as Signal 1 through switching on the coupling light, thus Bob had a two-photon polarization entangled state. The next work done by Bob should be decoding information by performing a Bell state measurement for a true implementation of QSDC. However due to the big difficulty in complete distinguishing four Bell states at present, the decoding process was not applied as the original protocol claimed in this experiment, we demonstrated the fact that Bob really obtained one of four Bell states corresponding to the encoding process through constructing the density matrix for every Bell state instead. This density matrix method is a standard method widely used for the verification of entanglement [30]. In addition, there exists a solution for implementing the complete and deterministic Bell state measurement based on linear optics, e.g. using hyper-entanglement [31, 32], therefore a complete demonstration of QSDC with quantum memories is possible in the future.

In the first round, we checked the entanglement without storage in MOT A/B and with no WP1 and WP2. This can be regarded as the first security check after the parties set up the quantum channel. This entanglement between Signal 1 and Signal 2 is written as $|\psi_3\rangle$, whose density matrix is illustrated in Figure 2. The fidelity of the state $|\psi_3\rangle$ was calculated by comparing it with the ideal density matrix, which is 93.1±1.0%, where $|\psi_3\rangle$ is

$$|\psi_3\rangle = |H_{S2}\rangle|H_{S1}\rangle + |V_{S2}\rangle|V_{S1}\rangle.$$

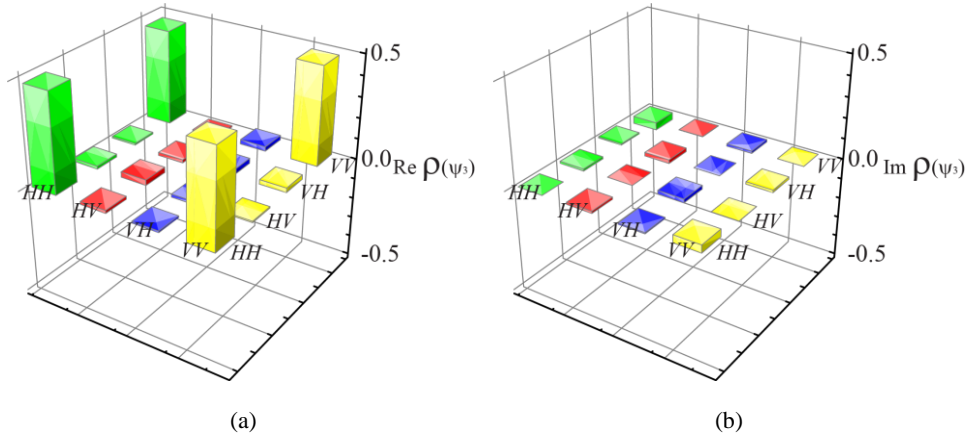

(a) (b)

Figure 2 Density matrix for $|\psi_3\rangle$. (a) and (b) are the real and imaginary parts, respectively.

Then, under the condition of storage in MOT A/B, and with no WP1 and WP2, this entanglement between the retrieved Signal 1 and Signal 2 is written as $|\psi_4\rangle$, whose density matrix is illustrated in Figure 3. State $|\psi_4\rangle$ has the same form as $|\psi_3\rangle$; that is, an encoding operation by unitary operation $I$, and the information value is "00," according to the previous agreement. The fidelity of $|\psi_4\rangle$ is 87.0±2.8% compared with the ideal density matrix.

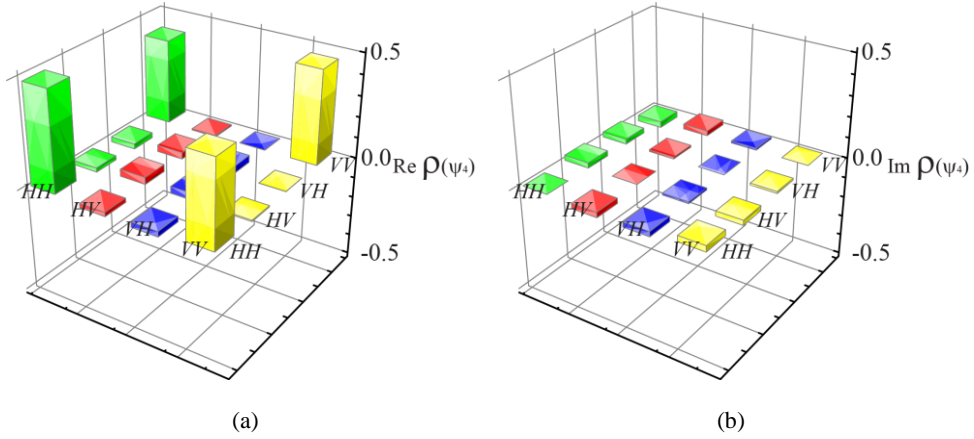

(a)                          (b)

Figure 3 Density matrix for $|\psi_4\rangle$. (a) and (b) are the real and imaginary parts.

In Figure 4, we also show the density matrices of the states $|\psi_5\rangle$ and $|\psi_6\rangle$ with a fidelity of 92.0±1.0% and 93.0±1.0% compared with the ideal density matrix, respectively. State $|\psi_5\rangle$ corresponds to the encoding operation $\sigma_z$ and the information value is "01," and state $|\psi_6\rangle$ corresponds to the encoding operation $\sigma_x$ and the information value is "10," where $|\psi_5\rangle$ and $|\psi_6\rangle$ can be written as

$$|\psi_5\rangle = |H_{S2}\rangle|H_{S1}\rangle - |V_{S2}\rangle|V_{S1}\rangle$$
$$|\psi_6\rangle = |H_{S2}\rangle|V_{S1}\rangle + |V_{S2}\rangle|H_{S1}\rangle.$$

To obtain $|\psi_5\rangle$, we set WP1's fast axis to angle $\theta_1=0$ with respect to the vertical axis at the same time as withdrawing WP2. To obtain $|\psi_6\rangle$, we set WP1's fast axis to angle $\theta_1=\pi/4$ with respect to the vertical axis at the same time as withdrawing WP2.

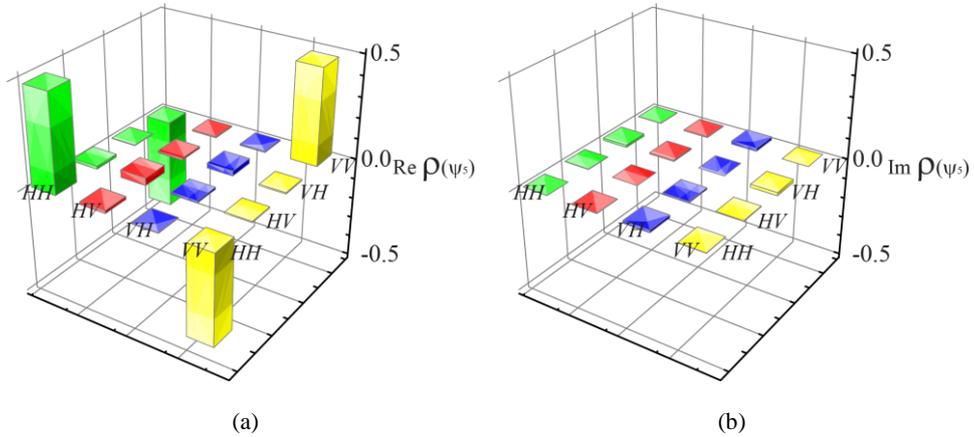

(a)                          (b)

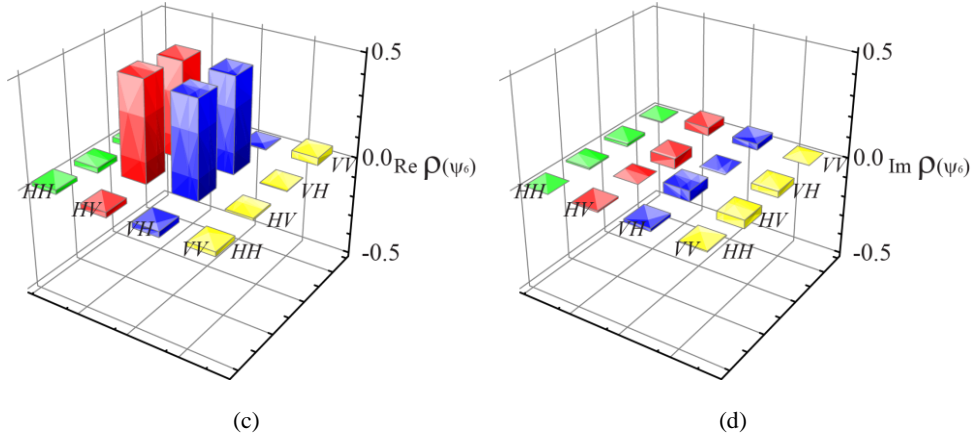

(c)                    (d)

Figure 4 Density matrix for $|\psi_5\rangle$ and $|\psi_6\rangle$. (a) and (b) are the real and imaginary parts, respectively, of the density matrix for $|\psi_5\rangle$, and (c) and (d) are the real and imaginary parts, respectively, of the density matrix for $|\psi_6\rangle$.

We further encode this entanglement through an operation with both WP1 and WP2 to set the fast axis of WP1 to angle $\theta_1=0$ with respect to the vertical axis and the fast axis of WP2 to angle $\theta_2=\pi/4$ with respect to the vertical axis. This entangled state can be expressed as $|\psi_7\rangle$, whose density matrix is illustrated in Figure 5 with a fidelity of 88.3±2.0% compared with the ideal density matrix. This state $|\psi_7\rangle$ corresponds to the encoding operation $\sigma_{iy}$, and the information value is "11":

$$|\psi_7\rangle = |V_{S2}\rangle|H_{S1}\rangle - |H_{S2}\rangle|V_{S1}\rangle.$$

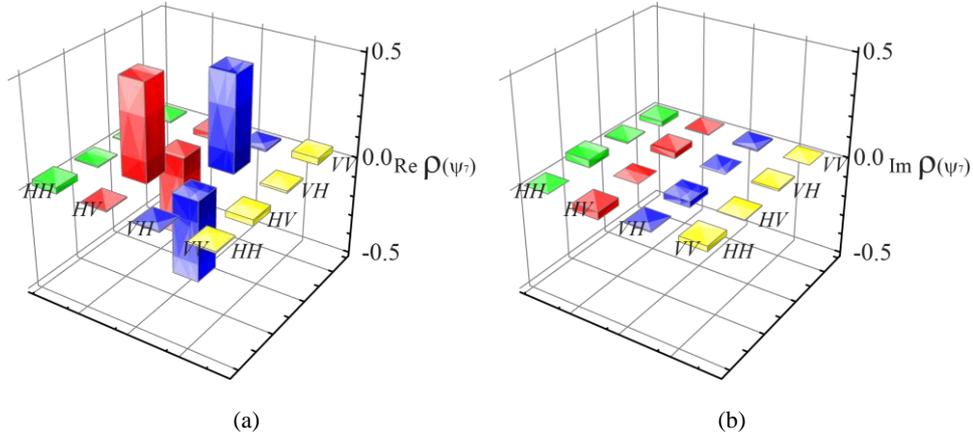

(a)                    (b)

Figure 5 Density matrix for $|\psi_7\rangle$. (a) and (b) are the real and imaginary parts, respectively.

QSDC protocols provide an efficient approach for sending information directly without sharing a key. Analogous with QKD, the security of the QSDC protocol relies on the basic principle of quantum mechanics, such as the uncertainty principle and no-cloning theorem.

In this experiment, we did not directly create the two-photon polarization Bell state, but a hybrid atom-photon entangled state $|\psi_1\rangle$ first, and then memory-memory entanglement $|\psi_2\rangle$. By exploiting the idea of dense coding, the QSDC can transmit information with a higher capacity than quantum teleportation. In recent years, quantum teleportation of multiple degrees of freedom of a single photon has been achieved both in theory and experiments [12, 33]; it requires the sophisticated quantum non-demolition measurement and complete hyper-entanglement Bell-state measurement, which is not an easy task in the current experiment.

We have reported the QSDC with quantum memory, thereby demonstrating the ability of secure direct communication. Regarding the coincidence rate, the bit rate is approximately 2.5/s and the error rate is approximately 0.1, which accounts for the 90% fidelity. The efficiency of direct communication is first limited by storage efficiency. In this experiment, the storage efficiency for Signal 1 in MOT B was approximately 25%. The efficiency of the quantum memory is mainly limited by the low OD. This efficiency can be increased to near-unity for coherent light with an OD up to 1,000 [34]. Using a backward direction of retrieval, we can overcome the limitation and achieve storage efficiency of more than 90% [35]. In the original QSDC protocol, to increase communication efficiency, Alice can prepare the ordered *N* pairs of the same Bell states and distribute the entangled states to Bob simultaneously, which is called "block transmission" technology. The quantum repeaters with multiplexed memory may be a good tool to achieve block transmission [36, 37]. Similar to quantum teleportation, QSDC depends on the distribution of the entanglement in distant locations. For long-distance QSDC, the large photon loss and decoherence in optical fibers requires the use of quantum repeaters, which is a substantial experimental challenge. In free space, photon loss and decoherence are almost negligible in the outer atmosphere. Free-space QSDC on the hundred-kilometer scale is possible based on the fact that experiments for quantum teleportation and entanglement distribution over 100 km have been well demonstrated [10, 11].

We have reported the first proof of the QSDC protocol with atomic quantum memory in principle. We demonstrated almost whole process of the QSDC protocol, including the generation of entanglement, channel security check, distribution of the entangled photons, storage, and encoding process with the decoding process replaced by constructing the density matrix. We want to mention a fact again that the decoding process is not applied as the original protocol claimed in the experiment due to the difficulty in distinguishing four Bell states, but the complete and determinate Bell state measurement for decoding can be achieved for example, by using hyper-entanglement [31, 32], therefore a complete demonstration of QSDC is feasible. We believe that our experiment will help the fundamental tests for future satellite-based ultra-long-distance and global QSDC, and quantum secure networks.

This work was supported by the National Natural Science Foundation of China (Grant Nos. 61275115, 61435011, 61525504, 11604322, 11474168 and 61401222).